\documentclass[twocolumn,twoside,slac_two]{revtex4}
\usepackage{graphicx}
\usepackage{fancyhdr}
\begin{document}

\title{$B \rightarrow K^{*} \gamma$ Decay within MSSM}
\author{Ciprian Dariescu and Marina-Aura Dariescu}
\affiliation{Department of Solid State and Theoretical Physics,
Al. I. Cuza University, Ia\c{s}i, Romania}

\begin{abstract}
The paper deals with a next-to-leading order analysis of the
radiative $B \to K^* \gamma$ decay. Working in a PQCD approach, we
compute the correction to the essential form factor, coming from a
single gluon exchange with the spectator. We investigate the
supersymmetry effects on the branching ratio and direct CP
asymmetry and constrain the squark mixing parameter $\left(
\delta_{23}^d \right)_{LR}$.
\end{abstract}

\maketitle

\thispagestyle{fancy}

\section{INTRODUCTION}

After the Cabibbo-favoured $b \to s \gamma$ mode was first
reported, in 1993, by CLEO II [1] and updated in 1995 [2], the
exclusive radiative decays, $B \to K^* \gamma$ and $B \to \rho
\gamma$, as well as the inclusive ones, $B \to X_{s(d)} \gamma$,
have become main targets for both experimental and theoretical
investigations. The exclusive modes, which are easier to be
experimentally investigated [3, 4, 5], but less theoretically
clear, have been worked out in different approaches. For example,
the spin symmetry for heavy quarks combined with wave function
models [6, 7] or the heavy quark effective theory when both $b$
and $s$ are heavy [8] have been used. Also, perturbative QCD
(PQCD) formalisms, introduced for exclusive nonleptonic
heavy-to-light transitions, have been extended to account for the
radiative decays. Recently, detailed analyses of $B \to K^*
\gamma$ and $B \to \rho \gamma$, in the next-to-leading order
(NLO), with the inclusion of hard spectator and vertex
corrections, have been performed [9-11] and a consistent
treatment, based on a new factorization formula, has been proposed
[12]. Besides an independent determination of the $| V_{td} /
V_{ts}|$ ratio, the $b \to s \gamma$ decays are suitable for
studying the viability of SUSY extensions of the SM, in view of
flavour changing neutral currents (FCNC) and CP tests, and for
imposing constraints on the supersymmetric benchmark scenarios
[13, 14].

The aim of the present paper is to analyse the $B \to K^* \gamma$
decay, in the minimal supersymmetric SM (MSSM) context. First, at
next-to-leading order, we compute the hard-spectator correction to
the essential form factor. In this respect, we employ the PQCD
approach developed by Szczepaniak {\it et al.} for decays
dominated by tree diagrams [15] and later extended to {\it
penguin} processes [16]. In order to fit the $Br$ estimation with
data and since large CP asymmetries, which are expected to be
measured more precisely in the near future, at the hadron
machines, seam to have a new physics origin, we extend our
analysis beyond the SM. In this respect, we make use of the mass
insertion method and include, in the Wilson coefficients
$C_{7,8}$, gluino-mediated FCNC contributions. Finally, the $Br$
data and the BaBar allowed range for direct CP asymmetry are used
to constrain the squark mixing parameter $\left( \delta_{23}^d
\right)_{LR}$.
\\
\section{NLO CORRECTION TO FORM FACTOR}

The effective Hamiltonian describing the $B \to K^* \gamma$
radiative decay is given by [9, 10]
\begin{equation}
H \, = \, \frac{G_{F}}{\sqrt{2}} \, \lambda_p \left[ C_{7} {\cal
O}_{7} + C_1 {\cal O}^p_1 + C_8 {\cal O}_8 \right] \, ,
\end{equation}
where $\lambda_p \equiv  V_{pb} V_{ps}^{*}$, with $p$ summed over
$u$ and $c$, and $C_1$, $C_{7}$, $C_8$ are the effective Wilson
coefficients at $\mu =m_b$. The hadronic matrix elements of the
four-fermion operator and of the electromagnetic and chromagnetic
penguin operators
\begin{eqnarray}
{\cal O}_1^p & = & \left( \bar{s} \gamma_{\mu} (1-\gamma_5 ) p
\right) \left( \bar{p} \gamma^{\mu} (1-\gamma_5 ) b \right) \, ,
\nonumber \\* {\cal O}_{7} & = & \frac{e m_b}{8 \pi^{2}} \, \left[
\bar{s} \sigma^{\mu \nu} (1+\gamma_{5}) b \right] F_{\mu \nu} \, ,
\nonumber \\* {\cal O}_{8} & = & \frac{g_s m_b}{8 \pi^{2}} \,
\left[ \bar{s} \sigma^{\mu \nu} (1+\gamma_{5}) T_i b \right]
G^i_{\mu \nu} \, ,
\end{eqnarray}
possess the general Lorentz decomposition:
\begin{widetext}
\begin{eqnarray}
& & \langle K^* | \bar{s} \gamma_{\mu} b | \bar{B} \rangle \, = \,
\frac{2 i \, V(q^2)}{m_B + m_{K^*}} \, \varepsilon_{\mu \nu \alpha
\beta} \epsilon^{* \nu} P_{K^*}^{\alpha} P_B^{\beta} \, ,
\nonumber \\* & & \langle K^* | \bar{s} \gamma_{\mu} \gamma_5 b |
\bar{B} \rangle \, = \, 2 m_{K^*} A_0 (q^2) \frac{(\epsilon^*
q)}{q^2} \, q_{\mu} + A_1 (q^2) (m_B + m_{K^*} ) \left[
\epsilon^*_{\mu} - \frac{(\epsilon^* q)}{q^2} \, q_{\mu} \right]
\nonumber \\* & & - A_2 (q^2) \, \frac{(\epsilon^* q)}{m_B +
m_{K^*}} \left[ ( P_B + P_{K^*})_{\mu} -
\frac{(m_B^2-m_{K^*}^2)}{q^2} \, q_{\mu} \right] , \; \; \; \; \;
\;
\\*
& & \langle K^* | \bar{s} \sigma_{\mu \nu} q^{\nu} b | \bar{B}
\rangle \, = \, 2 \, T_1(q^2) \, \varepsilon_{\mu \nu \alpha
\beta} \epsilon^{* \nu} P_{K^*}^{\beta} P_{B}^{\alpha} \, ,
\nonumber \\* & &  \langle K^* | \bar{s} \sigma_{\mu \nu} \gamma_5
q^{\nu} b | \bar{B} \rangle \, = \, - \, i \, T_2(q^2) \left[
\left( m_B^2 - m_{K^*}^2 \right) \epsilon^{*}_{\mu} - ( \epsilon^*
q) \left( P_B + P_{K^*} \right)_{\mu} \right] \nonumber \\* & & -
i \, T_3 (q^2) ( \epsilon^* q) \left[ q_{\mu} - \frac{q^2}{m_B^2
-m_{K^*}^2} \, \left( P_B + P_{K^*} \right)_{\mu} \right],
\end{eqnarray}
\end{widetext}
where $q_{\mu}$ is the the momentum of the photon and
$\epsilon^{\nu}$ is the $K^*$ 4-vector polarization. The form
factors are not known from first principles and this imprecise
knowledge is a major source of mismatch between theory and data as
well as between different theoretical estimations [9].

In the heavy quark limit, $m_b \gg \Lambda_{QCD}$, by neglecting
the corrections of order $1/m_b$ and $\alpha_s$, one has the
following relation among the form factors [9]
\begin{eqnarray}
& & \frac{m_B}{m_B + m_{K^*}} \, V(0) = \frac{m_B + m_{K^*}}{m_B}
\, A_1(0) \nonumber \\* & & = T_1(0) = T_2(0) \equiv F_{K^*}(0)
\end{eqnarray}
This is broken when one includes QCD radiative corrections coming
from vertex renormalization and hard gluon exchange with the
spectator. At order $\alpha_s$, the form factors encode strong
interaction effects by receiving an additive correction from hard
spectator interaction [11]. We recommend [9, 10, 12] for detailed
analyses of both factorizable and nonfactorizable vertex and
hard-spectator contributions, involving the operators ${\cal
O}_7$, ${\cal O}_8$ and penguin-type diagrams of ${\cal O}_1$.
However, it has been stated that factorization holds, at large
recoil and leading order in $1/m_b$, and quantitative tests for
proving QCD factorization at the level of power corrections have
been provided [17].

For a consistent treatment of radiative decays, at next-to-leading
order in QCD, a novel factorization formula have been proposed in
[12]. In this approach, the hadronic matrix elements in (1) are
written in terms of the essential form factor, which describes the
long-distance dynamics and is a nonperturbative object, and of the
hard-scattering kernels, $T_i^I$ and $T_i^{II}$, including the
perturbative short-distance interactions, as
\begin{equation}
\langle K^* \gamma | {\cal O}_i | \bar{B} \rangle = \left[ F_{K^*}
(0) T_i^I + \phi_B \otimes T_i^{II} \otimes \phi_{K^*} \right]
\cdot \eta \, ,
\end{equation}
where $\eta$ is the photon polarization. When the dominant
contribution comes from ${\cal O}_7$, we use (4) to write down the
decay amplitude as
\begin{widetext}
\begin{equation}
{\cal A}^{(0)} \, = \, \frac{G_F}{\sqrt{2}} \, \lambda_p \,
\frac{e m_b(\mu)}{2 \pi^2} \, C_7 (\mu) \, F_{K^*}(0) \left[
\varepsilon_{\mu \nu \alpha \beta} \eta^{\mu} \epsilon^{* \nu}
P_{K^*}^{\alpha} P_B^{\beta} - i \, (P_{K^*} q) (\eta \epsilon^* )
+ i  ( \epsilon^* q) (\eta P_{K^*} ) \right] ,
\end{equation}
and consequently the branching ratio reads
\begin{equation}
Br^{LO} = \tau_B \, \frac{G_F^2 \alpha |\lambda_p|^2 \, m_b^2}{32
\pi^4} \, m_B^3 \, ( 1-z^2)^3 | C_7 (m_b) |^2 |F_{K^*} (0)|^2 \, ,
\end{equation}
\end{widetext}
where $z=m_{K^*} /m_B$. At next-to-leading order in $\alpha_s$,
one has to consider, in (6), the contributions to the hard
scattering kernels $T^I_i$ coming from the operators ${\cal O}_1$
and ${\cal O}_8$. These have been evaluated in [12] and bring (7)
to the expression
\begin{widetext}
\begin{equation}
{\cal A} = \frac{G_F}{\sqrt{2}} \, \lambda_p \, \frac{e
m_b(\mu)}{2\pi^2} \left[ C_7 + \frac{\alpha_s C_F}{4 \pi} \left(
C_1 G_1^p + C_8 G_8 \right) \right] F_{K^*}(0) \left[
\varepsilon_{\mu \nu \alpha \beta} \eta^{\mu} \epsilon^{* \nu}
P_{K^*}^{\alpha} P_B^{\beta} - i \, (P_{K^*} q) (\eta \epsilon^* )
+ i  ( \epsilon^* q) (\eta P_{K^*} ) \right] ,
\end{equation}
where $C_F = (N^2-1)/(2N)$, $N=3$, and
\begin{eqnarray}
G_1 (s) & = & - \, \frac{833}{162} \, - \, \frac{20 \, i \pi}{27}
+ \frac{8 \pi^2}{9} \, s^{3/2} + \, \frac{2}{9} \left[ 48 + 30 i
\pi - 5 \pi^2 - 2 i \pi^3 - 36 \zeta (3) + (36+6i \pi -9 \pi^2 )
\ln s \right. \nonumber \\* & & + \left. (3+6i \pi ) \ln^2 s +
\ln^3 s \right] s  + \, \frac{2}{9} \left[ 18 + 2 \pi^2 -2i \pi^3
+ (12 -6 \pi^2 ) \ln s + 6i \pi \ln^2 s + \ln^3 s \right] s^2
\nonumber \\* & & + \, \frac{1}{27} \left[ -9 +112 i \pi -14 \pi^2
+ (182-48i \pi ) \ln s - 126 \ln^2 s \right] s^3 \, , \nonumber
\\* G_8 & = & \frac{11}{3} - \frac{2 \pi^2}{9} + \frac{2i \pi}{3}
\, ,
\end{eqnarray}
\end{widetext} with $s_c = m_c^2/m_b^2$ and $\mu = m_b$.

Going further, we add factorizable NLO hard-spectator corrections,
to the form factor $F_{K^*} (0)$. For a single gluon exchanged
with the spectator (see Fig. 1), we extend the PQCD approach,
developed by Szczepaniak {\it et al.} for {\it heavy-to-light}
transitions dominated by tree diagrams [15], to the so-called {\it
penguin} processes.

\begin{figure}
\includegraphics{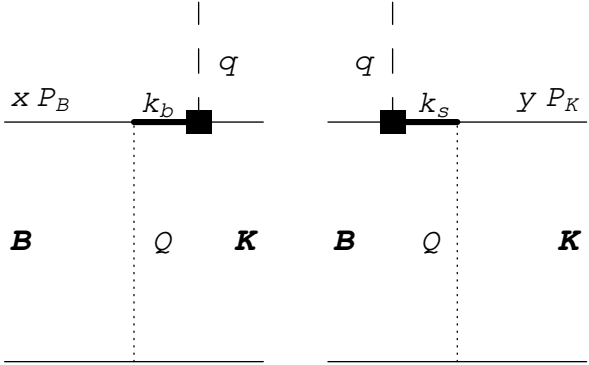}
\caption{The Feynman contributing diagrams in the hard scattering
amplitude $T_{\mu}$. The gluon and photon are respectively
represented by dotted and dashed lines.}
\end{figure}

For the operator ${\cal O}_7$, one has to evaluate the following
trace over spin, flavor and color indices, and integrate over
momentum fractions [16]
\begin{eqnarray}
T_{\mu} & = & {\rm Tr} \left[\bar{\phi}_{K^*} \sigma_{\mu \nu} (1
+ \gamma_{5}) q^{\nu} \frac{\rlap{/}{k}_{b} + m_{b}}{k_{b}^{2} -
m_{b}^{2}} \gamma_{\alpha} \phi_{B} \gamma^{\alpha} \frac{4
g_{s}^{2}}{Q^{2}} \right] \nonumber \\* & + & \,  {\rm Tr}
\left[\bar{\phi}_{K^*} \gamma_{\alpha} \frac{\rlap{/}
k_s}{k_{s}^{2}} \sigma_{\mu \nu} (1 + \gamma_{5}) q^{\nu} \phi_{B}
\gamma^{\alpha} \frac{4 g_{s}^{2}}{Q^{2}} \right] ,
\end{eqnarray}
where $Q^2 \approx - (1-x)(1-y)m_B^2$. The $B$ meson wave function
\begin{equation}
\phi_{B} = \frac{f_B}{12} \, \varphi_{B}(x)(\rlap{/}{P}_{B} +
m_{B}) \gamma_{5}
\end{equation}
contains a strongly peaked distribution amplitude, around $a
=\lambda_B/m_B \approx 0.072$, for $\lambda_B =0.38$. The $K^*$ is
described by the wave function
\begin{equation}
\phi_{K^*} = \frac{f_{K^*}}{12} \, \varphi_{K^*}(y)
\rlap{/}{P}_{K^*} \rlap{/}{\epsilon} \, ,
\end{equation}
where the light-cone distribution amplitude, $\varphi_{K^*}(y)$,
has the following expansion in Gegenbauer polynomials [18]
\begin{eqnarray}
\varphi_{K^*} (y) & = & 6y(1-y)[ 1+ \alpha_1^{K^*}
C^{(3/2)}_1(2y-1) \nonumber \\* & + & \alpha_2^{K^*}
C^{3/2}_2(2y-1) + ...],
\end{eqnarray}
with $C^{3/2}_1(u) = 3 u$, $C^{3/2}_2(u) = (3/2)(5u^2-1)$,
$\alpha_1^{K^*} (m_b) = 0.18 \pm 0.05$, and $\alpha_2^{K^*} (m_b)
= 0.03 \pm 0.03$. Performing the calculations in (11) and using
the form factors decomposition (4), we identify the spectator
contribution to the essential form factor as
\begin{equation}
F^{sp} (a) \, = \, \frac{g_{s}^{2}}{9} \frac{f_B f_{K^*}}{m_B
\lambda_B } \int_0^{1-a} dy \; \frac{(2-y)}{(1-y)^2} \,
\varphi_{K^*} (y) \, ,
\end{equation}
where the $K^*$-mass has been neglected. Since we have introduced
a cut-off in $y \to 1$, the form factor correction (15) depends on
the peaking parameter $a$ and this is a main uncertainty in our
calculations. For the following input values: $\alpha_s (\mu =
Q^2) \approx 0.38$, $f_B = 0.180$ GeV, $f_{K^*}^{\perp} = 0.185$
GeV and $a=0.072$, we get
\begin{equation}
F^{sp} (0.072) \, = \, 0.1475
\end{equation}
Introducing, in (9), the total form factor $F_{K^*}(0) + F^{sp}
(0.072) = 0.38 + 0.1475$, where we have used the prediction from
the QCD sum rules analysis $F_{K^*} (0) = 0.38 \pm 0.06$, the
branching ratio gets significantly enhanced to the value $Br^{NLO}
= 6.97 \times 10^{-5}$. This is comparable to the average
theoretical prediction $\left( 7.5 \pm 0.3 \right) \times 10^{-5}$
[9, 11, 12], but is above the experimental data:
\begin{widetext}
\begin{eqnarray}
Br(B^+ \to K^{*+} \gamma) = \left \lbrace
\begin{array}{lc}
\left( 3.83 \pm 0.62 \pm 0.22 \right) \times 10^{-5}
& ({\rm BaBar} \; [3]) \\
\left(  3.76_{-0.83}^{+0.89} \pm 0.28
\right) \times 10^{-5} & ( {\rm CLEO} \; [4]) \\
\left( 3.89 \pm 0.93 \pm 0.41 \right) \times 10^{-5} & ({\rm
Belle} \; [5])
\end{array}
\right. \nonumber \\* Br(B^0 \to K^{*0} \gamma) = \left \lbrace
\begin{array}{lc}
\left( 4.23 \pm 0.40 \pm 0.22 \right) \times 10^{-5}
& ({\rm BaBar} \; [3]) \\
\left( 4.55^{+0.72}_{-0.68} \pm 0.34
\right) \times 10^{-5} & ({\rm CLEO} [4]) \\
\left( 4.96 \pm 0.67 \pm 0.45 \right) \times 10^{-5} & ({\rm
Belle} [5])
\end{array}
\right. \nonumber
\end{eqnarray}
\end{widetext} whose world average value, over the $B^{\pm}$ and
$B^0$ decay modes, is
\begin{equation}
Br_{exp} (B^{\pm} \to K^{*\pm} \gamma ) \, = \, \left( 4.22 \pm
0.28 \right) \times 10^{-5}
\end{equation}

For the direct CP asymmetry,
\begin{equation}
a_{CP} \, = \, \frac{\Gamma(\bar{B} \to K^* \gamma ) - \Gamma ( B
\to K^* \gamma )}{ \Gamma(\bar{B} \to K^* \gamma ) + \Gamma ( B
\to K^* \gamma )} \, ,
\end{equation}
which is predicted by the SM to be $a_{CP} < | 0.005 |$, the BaBar
and CLEO data are: $a_{CP} = - 0.044 \pm 0.076 \pm 0.012$ (BaBar
[3]) and $a_{CP} = 0.08 \pm 0.13 \pm 0.03$ (CLEO [4], for the sum
of neutral and charged $B \to K^* \gamma$ decays). Even though the
data look consistent with the SM result within $1 \sigma$, it is
too early to draw a definitive conclusion because of the large
errors. With BaBar and Belle large data samples, there is hope for
precise measurements of large CP asymmetries and this makes room
for new physics. Moreover, it has been stated that CP asymmetries
larger than few percent would have a dominantly supersymmetric
origin [19].

\section{Branching ratio and direct CP
asymmetry within MSSM}

Following this idea, let us analyse the $B \to K^* \gamma$ decay
in the MSSM context. Using the mass insertion approximation, [20],
we incorporate, in the Wilson coefficients $C_{7}$ and $C_8$, the
FCNC SUSY contributions
\begin{widetext}
\begin{eqnarray}
C_{7}^{SUSY} (M_{SUSY})  & = & \frac{\sqrt{2} \pi \alpha_s}{G_F
(V_{ub} V_{us}^* + V_{cb} V_{cs}^*) m_{\tilde{g}}^2} \, \left(
\delta_{23}^d \right)_{LR} \frac{m_{\tilde{g}}}{m_b} \, F_0(x) \,
; \nonumber \\* C_{8}^{SUSY} (M_{SUSY}) & = & \frac{\sqrt{2} \pi
\alpha_s}{G_F (V_{ub} V_{us}^* + V_{cb} V_{cs}^*) m_{\tilde{g}}^2}
\left( \delta_{23}^d \right)_{LR} \frac{m_{\tilde{g}}}{m_b} \,
G_0(x) \, ,
\end{eqnarray}
where
\begin{eqnarray}
F_0 (x) & = & - \; \frac{4x}{9(1-x)^4} \, \left[ 1+4x-5x^2+4 x
\ln(x) + 2 x^2 \ln (x) \right] , \nonumber \\* G_0 (x) & = &
\frac{x}{3(1-x)^4} \, \left[ 22-20x-2x^2+16 x \ln(x) -x^2 \ln (x)
+ 9 \ln (x) \right] \; \; \; \;
\end{eqnarray}
\end{widetext}
In (20), $x=m_{\tilde{g}}^2 / m_{\tilde{q}}^2$ is expressed in
terms of the gluino mass, $m_{\tilde{g}}$, and an average squark
mass, $m_{\tilde{q}}$. We underline that, in the expressions of
$C_{7,8}^{SUSY} (M_{SUSY})$, we have kept only the left-right
squark mixing parameter $\left( \delta_{23}^d \right)_{LR} \, = \,
\left( \Delta_{bs} \right)/ m_{\tilde{q}}^2$ since, being
proportional to the large factor $m_{\tilde{g}} /m_b$, it has a
significant numerical impact on the branching ratio value. The
quantities $\Delta_{bs}$ are the off-diagonal terms in the
sfermion mass matrices, connecting the flavours $b$ and $s$ along
the sfermion propagators. In these assumptions, the total Wilson
coefficients, encoding the new physics, become
\begin{eqnarray}
C_{7}^{total} [x, \delta] & = & C_{7} (m_b) + C_{7}^{SUSY} (m_b)
\; , \nonumber \\* C_{8}^{total} [x, \delta] & = & C_{8} (m_b) +
C_{8}^{SUSY} (m_b) \; ,
\end{eqnarray}
where $C^{SUSY}_{7,8} (m_b)$ have been evolved from $M_{SUSY} =
m_{\tilde{g}}$ down to the $\mu =m_b$ scale, using the relations
[21]
\begin{eqnarray}
C_{8}^{SUSY} (m_b) & = & \eta C_{8}^{SUSY}(m_{\tilde{g}} ) \, ,
\nonumber \\* C_{7}^{SUSY} (m_b) & = & \eta^2
C_{7}^{SUSY}(m_{\tilde{g}} ) \nonumber \\* & + & \frac{8}{3} (\eta
- \eta^2) C_{8}^{SUSY}(m_{\tilde{g}} ) \, ,
\end{eqnarray}
with
\begin{equation}
\eta \, = \, \left( \alpha_s(m_{\tilde{g}})/ \alpha_s(m_t)
\right)^{2/21} \left( \alpha_s(m_t)/ \alpha_s(m_b) \right)^{2/23}
\end{equation}

Finally, putting everything together, we replace, in (9), the
Wilson coefficients $C_7$ and $C_8$ respectively by $C_7^{total}[x
, \delta]$ and $C_8^{total} [x , \delta ]$, the form factor
$F^{K^*} (0)$ by $F^{K^*} (0) + F^{sp} (a)$ and, consequently, the
branching ratio is
\begin{widetext}
\begin{eqnarray}
Br^{total} & = & Br^{SM + SUSY} \, = \, \tau_B \, \frac{G_F^2
\alpha m_b^2}{32 \pi^4} \, m_B^3 \, ( 1-z^2)^3 \; |F_{K^*} (0)
+F^{sp} (a)|^2 \nonumber \\* & \times & \left| \lambda_p \left[
C_7^{total} [x, \delta ] + \frac{\alpha_s C_F}{4 \pi} \left( C_1
G_1^p + C_8^{total} [x, \delta ] G_8 \right) \right] \right|^2
\end{eqnarray}
\end{widetext} For a given $x$ and $\delta \equiv \rho e^{i
\varphi}$, the total branching ratio (24) is depending on three
free parameters: $a , \, \rho , \, \varphi$. One can notice that,
by including the hard gluon contributions as a correction to
$F_{K^*} (0)$, the direct CP asymmetry parameter, (18), is free of
the uncertainties coming from the form factors.

In what it concerns the gluino, as its pair production cross
section has large cancellations in the $e^+ e^-$ annihilation,
there is hope that the laser-backscattering photons will provide a
precise gluino mass determination [22]. For a wide range of squark
masses, a gluino mass of 540 GeV may be measured, with a precision
of at least $\pm 2 \dots 5$, at the multi-TeV linear collider at
CERN.

In the next coming discussion, we use the following input
parameters: $m_b (m_b)=4.2$ GeV, $\alpha =1/137$,
$|V_{tb}V_{ts}^*| = 0.0396 \pm 0.002$, $\tau_{B^0} = \left( 1.546
\pm 0.018 \right)$ ps, $m_{\tilde{q}} = 500$ GeV and $x$ has the
values $x_l =0.3$, $x_0 =(540/500)^2$ and $x_g =3$ (where $l(g)$
comes from $m_{\tilde{g}}$ less (greater) than $m_{\tilde{q}}$).

In figure 2, we represent the contour plots on which the
$Br^{total}$, (24), is equal to the world average data (17) and
the BaBar constraint [3]
\begin{equation}
- \, 0.17 < a_{CP} < 0.082
\end{equation}
When $\left \lbrace \rho , \, \varphi \right \rbrace \in [0, \,
0.03] \times [ -  \pi /2 , \, \pi /2 ]$, we get for (17) three
dashed lines, with increasing thickness, as $x$ goes from $x_l$ to
$x_g$. Correspondingly, for $a_{CP}$, we get three pairs of solid
curves: the lower ones, for $a_{CP} = - 0.17$, and the upper ones,
for $a_{CP} =0.082$. These solid contours close inside the values
of direct CP asymmetry which do not agree with (25). One is able
to constrain $\left( \delta_{23}^d \right)_{LR}$ to the segments
of the $Br$-plots outside the solid contours, for each $x$. For
example, when $m_{\tilde{g}} > m_{\tilde{q}}$, all negative
phases, with suitable $\rho$'s, can accommodate both (25) and
(17). Also, $\rho$ is constrained by the experimental data on the
branching ratio of $B \to X_s \gamma$ to $\rho \leq 0.016$ [20,
23].

In figure 3, we represent the $Br^{total}$ (in units of $10^{-5}$)
and $a_{CP}$ (in units of $0.1$), with respectively dashed and
solid lines, as functions of $\varphi$, for $x=x_0$. As $\rho$
takes the following values: $\rho \in \left \lbrace 0.005 , \,
0.01 , \, 0.012 \right \rbrace$, we get three pairs of curves,
with increasing thickness. The horizontal dashed band corresponds
to the data $3.76 \times 10^{-5} < Br_{exp} < 4.68 \times 10^{-5}$
(in $90 \%$ C.L.), while the horizontal solid band stands for the
constraint (25). We notice that one should avoid the region
$\varphi \in \left[ - \pi/16 , \, \pi/16 \right]$ on the $\rho
\approx 0.01$ plot, since the CP asymmetry parameter drops
quickly, from positive to negative values much outside the
constraint (24).

\begin{figure}
\includegraphics[width=60mm]{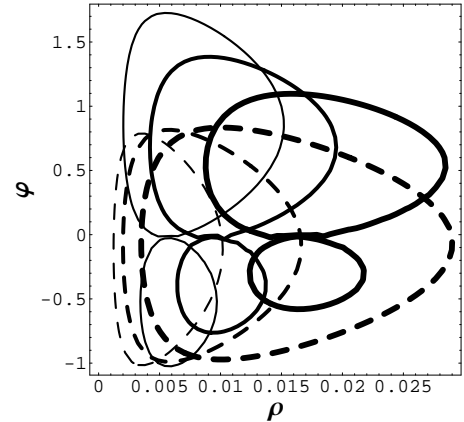}
\caption{Contour plots of total branching ratio, in units of
$10^{-5}$, fitting the world average data, (17), (the dashed
lines) and the BaBar constraint (25) (the solid lines), as
functions of $\rho$ and $\varphi$. The thickness of contours is
increasing as $x$ is taking the values: $x = \left \lbrace 0.3 ,
\, 1.16 , \, 3 \right \rbrace$. The solid curves close inside the
values of $a_{CP}$ which disagree with the constraint (25).}
\end{figure}

\begin{figure}
\includegraphics{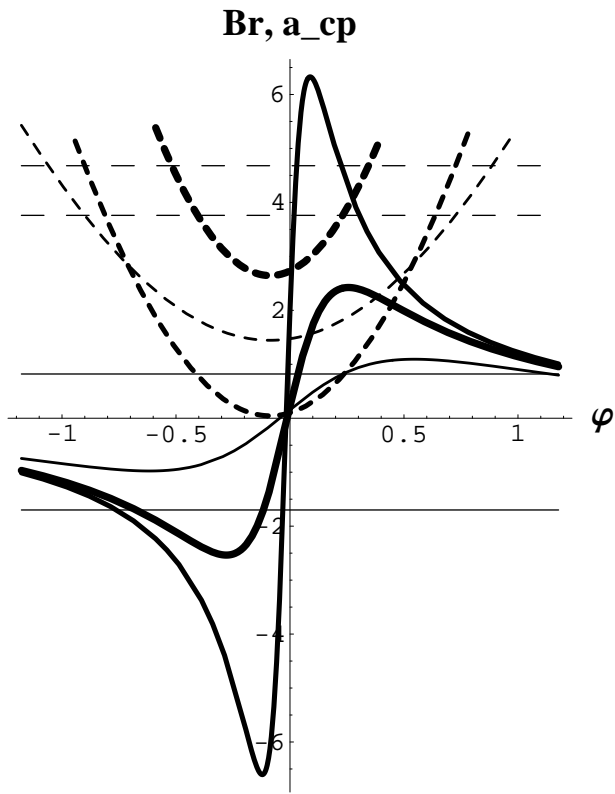}
\caption{The total branching ratio, in units of $10^{-5}$, (the
dashed lines) and $10 \times a_{CP}$ (the solid lines), as
functions of $\varphi$, for $x=x_0$. The thickness of plots
increases as $\rho$ is: $\rho = \left \lbrace 0.005 , \, 0.01 , \,
0.012 \right \rbrace$. The horizontal dashed band corresponds to
the world average branching ratio with $90 \%$ C.L. and the
horizontal solid one is for the constraint (25).}
\end{figure}

Finally, let us perform a numerical analysis, for $x=x_0$, and
increasing $\rho$, starting with $\rho =0.005$. As $\varphi \in
\left[ - 8 \pi/16 , \, - 4 \pi/16 \right] \cup \left[ 3 \pi/16 ,
\, 7 \pi/16 \right]$, the $Br^{total}$ and the direct CP asymmetry
are inside the ranges $10^5 \times Br^{total} \in \left[ 8.3 , \,
3.3 \right]$ and $\left[ 3.1 , \, 8.1 \right]$, accommodating data
and other theoretical models predictions, and, respectively,
$a_{CP} \in \left[ -0.054 , \, - 0.093 \right] \cup \left[ 0.096 ,
\, 0.062 \right]$. When $\rho$ goes to bigger values, the two
$\varphi$ ranges, constrained by the allowed branching ratios, get
closer and $a_{CP}$ moves toward much bigger values. For example,
for $\rho = 0.01$ and $\varphi \in \left[ - 6 \pi/16 , \, - 4
\pi/16 \right] \cup \left[ 3 \pi/16 , \, 5 \pi/16 \right]$, one
gets $10^5 \times Br^{total} \in \left[ 8.1 , \, 3.6 \right]$ and
$\left[ 3.2 , \, 7.62 \right]$ and, respectively, $a_{CP} \in
\left[ -0.1 , \, - 0.16 \right] \cup \left[ 0.21 , \, 0.12
\right]$. We notice that only the negative $\varphi$-values lead
to $a_{CP}$ inside the BaBar constraint (25). For $\rho$ close to
the upper limit, $\rho = 0.015$, and $\varphi \in \left[ - 4
\pi/16 , \, - 2 \pi/16 \right] \cup \left[ \pi/16 , \, 3 \pi/16
\right]$, the values $10^5 \times Br^{total} \in \left[ 7.8 , \,
3.6 \right]$ and $\left[ 3.3 , \, 7.3 \right]$ are compatible with
a measurable $a_{CP} \approx \pm 0.12$.

\section{Conclusions}

In the present paper, we have analysed the radiative $B \to K^*
\gamma$ decay in the MSSM framework. First, we have used the PQCD
approach, developed by Szczepaniak {\it et al.} [15] and extended
to ^^ ^^ penguin'' processes [16], to compute the hard-spectator
contribution, $F^{sp} (a)$, to the essential form factor $F_{K^*}
(0)$. For the peaking parameter in the $B$ wave function $a=0.072$
and $F_{K^*} (0) =0.38$, the branching ratio becomes $Br^{NLO} =
6.97 \times 10^{-5}$, which is above the experimental data. In
order to reach an agreement with data and find large CP
asymmetries, which hopefully will be measured with a better
precision in the near future, we extend our analyses by including,
in the Wilson coefficients $C_{7,8}$, the SUSY contributions
coming from squark mixing parameter $\left( \delta_{23}^d
\right)_{LR} = \rho e^{i \varphi}$. Consequently, the total
branching ratio depends, besides $a$, on three (SUSY) parameters:
$x , \rho , \varphi$, while $a_{CP}$ is free of the uncertainties
coming from the corrected form factor. Using the graphs displayed
in figures 2 and 3, one is able to find ranges for the mass
insertion parameter $\left( \delta_{23}^d \right)_{LR}$. As an
example, for $x= (540/500)^2$, the world average branching ratio,
$Br_{exp} = 4.22 \times 10^{-5}$, can be accommodated for $\lbrace
\rho , \, \varphi \rbrace = \left \lbrace 0.005 , \, - \, \frac{4
\pi}{13} \right \rbrace$ or $\left \lbrace 0.01 , \, - \, \frac{4
\pi}{15} \right \rbrace$. The corresponding asymmetries, $a_{CP}
=-0.085$ and respectively $a_{CP} = -0.147$, are inside the BaBar
constraint (25).
\begin{acknowledgments}
The authors gratefully acknowledge the kind hospitality and
fertile environment of the University of Oregon where this work
has been carried out. Special thanks go to Damir Becirevic and
Vladimir Braun for useful discussions and fruitful suggestions.
This work is supported by the CNCSIS Type A Grant, Code 1160.
\end{acknowledgments}

\end{document}